\documentclass[prb,
twocolumn,
superscriptaddress,showpacs,amsmath,amssymb]{revtex4}

\begin{document}

\title{Varying Newton constant and black hole to white hole quantum tunneling}

\author{G.E.~Volovik}
\affiliation{Low Temperature Laboratory, Aalto University,  P.O. Box 15100, FI-00076 Aalto, Finland}
\affiliation{Landau Institute for Theoretical Physics, acad. Semyonov av., 1a, 142432,
Chernogolovka, Russia}

\date{\today}

\begin{abstract}
The thermodynamics of black holes is discussed for the case, when the Newton constant $G$ is not a constant, but is the thermodynamic variable. This gives for the first law of the Schwarzschild black hole thermodynamics: $dS_\text{BH}= -AdK + \frac{dM}{T_\text{BH}}$, where the gravitational coupling $K=1/4G$, $M$ is the black hole mass, $A$ is the area of horizon, and $T_\text{BH}$ is Hawking temperature.
From this first law it follows that the dimensionless quantity  $M^2/K$ is  the adiabatic invariant, which in principle can be quantized if to follow the Bekenstein conjecture. From the Euclidean action for the black hole it follows that $K$ and $A$ serve as dynamically  conjugate variables. This allows us to  calculate the quantum tunneling from the black hole to the white hole, and determine the temperature and entropy of the white hole.
\end{abstract}
\pacs{
}

\maketitle

\section{Introduction}

Bekenstein \cite{Bekenstein1974}  proposed 
that the horizon area $A$ is an adiabatic invariant and thus can be quantized according to the
Ehrenfest principle that classical adiabatic invariants may
correspond to observables with discrete spectrum.
Here we consider the black hole thermodynamics in case when the gravitational coupling $K$ is the variable thermodynamic quantity. The variable $K$ modifies the first law of the black hole thermodynamics and leads to the alternative adiabatic invariant, which is dimensionless and thus in principle can be quantized.

On the quantum level, the gravitational coupling $K$ becomes the variable, which is dynamically conjugate to the black hole area $A$. This allows us to study the black hole to white hole transition using the semiclassical description of the quantum tunneling and considering the trajectory in the $(K,A)$ phase space, which connects the black and white holes.
We also consider the temperature and entropy of the white hole, which is formed from the black hole by quantum tunneling and find that both the temperature and entropy of the white hole are negative.

\section{Modified first law of black hole thermodynamics }
\label{FirstLaw}

The Einstein-Hilbert gravitational action  is
\begin{equation}
S_\text{grav}= 
 \frac{1}{4\pi }\int d^3x dt\sqrt{-g} K{\cal R}  \,,
\label{eq:gravity_action}
\end{equation}
where ${\cal R}$ is the scalar curvature, and we choose the gravitational coupling   $K=1/(4G)=1/(4 l^2_{\rm Planck})$.
In the modified gravity theories, such as the scalar-tensor and $f(R)$ theories 
(see e.g. \cite{Starobinsky1980} and the latest paper \cite{Barrow2020} with references therein), the effective Newton ``constant`` $G$ can be space-time dependent, and thus is not fundamental.
Also it was suggested that there is
 connection  between  the gravitational coupling $K$ and the fine structure constant, see Refs. \cite{Landau1955,Akama,AkamaTerazawa,Terazawa,TerazawaAkama,Terazawa1981,KlinkhamerVolovik2005}. 

Here we assume that the variables, which enter the Einstein action -- the scalar  Riemann curvature $R$  and the gravitational coupling constant $K$ -- are the local thermodynamic variables, which are similar to temperature, pressure, chemical potential, number density, etc.  The Riemann
curvature as the covariant quantity may serve as one of the thermodynamical characteristics of the macroscopic matter \cite{Pronin1987}.  If so,  the gravitational coupling constant $K$ in front of the scalar curvature  in Einstein action also becomes the thermodynamic quantity, see Ref. \cite{KlinkhamerVolovik2008e}. The application of  this thermodynamics to the global object, such as black hole,  suggests that the variable $K$ (actually its asymptotic value at infinity) should enter the thermodynamic laws for the black hole. The corresponding thermodynamical conjugate to the global $K$ is obtained by the volume  integral of the  Riemann
curvature.

 In terms of this coupling $K$
the Hawking temperature of Schwarzschild black hole and its Bekenstein entropy are:
 \begin{equation}
  T_\text{BH}=\frac{K}{2\pi M}~~,~~S_\text{BH}
             = \frac{\pi M^2}{K}   \,.
\label{eq:HawkingT}
\end{equation}
 Then, using  the black hole area $A= \pi M^2/K^2$; the  gravitational coupling $K=1/4G$;  
the Hawking temperature $T_\text{BH}=M/2AK= K/2\pi M$; and the black hole entropy $S_\text{BH}=AK$, one obtains:
\begin{eqnarray}
dS_\text{BH}=d(AK)= \pi d (M^2/K)  =
\nonumber
\\
=- \pi \frac{M^2}{K^2}dK + 2\pi \frac{M}{K} dM \,.
\label{dS2}
\end{eqnarray}

This suggests the following modification of the first law of black hole thermodynamics in case if $K$ is a thermodynamic variable:
\begin{equation}
dS_\text{BH}= -AdK + \frac{dM}{T_\text{BH}}\,.
\label{dS}
\end{equation}
This modification is similar to the modification in terms of the moduli fields \cite{Gibbons1996}. But in our case  the thermodynamic variable, which is  conjugate to the thermodynamic variable $K$, is the product of the black hole area and the black hole temperature, $AT_\text{BH}$. On the other hand in dynamics, $K$ and $A$ are canonically conjugate, see Sec.\ref{Canonical}. 

In general, the variable $K$ is local and depends on space coordinate, but in the same way as for the moduli fields \cite{Gibbons1996} the black hole thermodynamics is determined by the asymptotic value of $K$ at spatial infinity. In Eq.(\ref{dS}) $K\equiv K({\infty})$ is the global quantity, which characterizes the quantum vacuum in full equilibrium, i.e. far from the black hole. The variable $K$ allows us to study the transition to the vacuum without gravity, i.e. to the vacuum where $K\rightarrow\infty$ and thus $G\rightarrow 0$, see Sec.\ref{Canonical}.

\section{Adiabatic change of $K$ and adiabatic invariant}
\label{adiabatic}

Let us change $K$ and $M$ adiabatically, i.e. at constant entropy. Then the equation $dS_\text{BH}=0$ gives 
\begin{equation}
\frac{dM}{dK}= AT_\text{BH}=\frac{M}{2K} \,.
\label{ratio}
\end{equation}
This shows that $M^2/K={\rm const}$ is the adiabatic invariant for the spherical neutral black hole, and thus according to  the Bekenstein conjecture \cite{Bekenstein1974}, it can be quantized in quantum mechanics:
\begin{equation}
\frac{M^2}{K}= a N \,.
\label{quantization}
\end{equation}
Here $N$ is integer, and $a$ is some fundamental dimensionless parameter of order unity.
If this conjecture is correct, for the entropy of Schwartzschild black hole one has
\begin{equation}
S_\text{BH}(N)=\pi \frac{M^2}{K}= \pi a N \,.
\label{quantizationSchwarz}
\end{equation}
Note that in this approach, the black hole area $A=S_\text{BH}/K= \pi M^2/K^2$ is not quantized, since it is not dimensionless, and thus the prefactor cannot be fundamental. On the other hand, the entropy is dimensionless and thus in principle can be expressed as some dimensionless function of integers.  

The Bekenstein idea on the role of adiabatic invariants in quantization of the black hole requires further consideration, see some approaches to that in Refs. \cite{Horowitz1996,Barvinsky2001,Barvinsky2002,Ansorg2012,Visser2012,Tharanath2013}.
We leave this problem for the future. This consideration should be supported by microscopic theory, see e.g. \cite{Carlip2014}.  The so-called $q$-theory can be exploited, which allows us to consider dark energy, dark matter, black holes and the varying gravitational coupling in the frame of the same effective theory of the quantum vacuum
\cite{KlinkhamerVolovik2008e,Klinkhamer2017,Klinkhamer2019}.

\section{$A$ and $K$ as canonically conjugate variables and black-hole -- white-hole quantum tunneling}
\label{Canonical}

Till now we discussed the black hole thermodynamics, where the adiabatic invariant $M^2/K$ arises. Now we move to the dynamics of black hole, and apply the varying $K$ approach to the consideration of the quantum tunneling of the black hole to white hole. The quantum tunneling can be considered in a semiclassical approximation in the same manner as Hawking radiation is discussed in terms of   semiclassical quantum tunneling \cite{Volovik1999,Wilczek2000,Akhmedov2006,Volovik2009,Vanzo2011}.
The quantum mechanical  treatment of the black hole can be obtained,
if one finds the relevant canonically conjugate variables describing the black hole dynamics. Such approach has been suggested in Ref. \cite{Vagenas2011}, where the  canonically conjugate variables have been determined in the Euclidean time.  Since in Euclidean time the action for the black hole is equal to its entropy, $I_\text{E}=S_\text{BH}$,
in the theory with varying gravitational coupling $K$,   the proper  canonically conjugate variables are the gravitational coupling $K$ and the black hole area $A$. This allows us to consider the quantum mechanical tunneling from the black hole to the white hole, which was discussed in Refs. \cite{Barcelo2014,Barcelo2017,Rovelli2018,Rovelli2019,Rovelli2018b,Uzan2020,Uzan2020b,Uzan2020c,Bodendorfer2019} and references therein. 

The process of quantum tunneling of macroscopic objects is well known in condensed matter physics, where the collective variables are used, which describe the collective dynamics of a macroscopic body \cite{LifshitzKagan1972,IordanskiiFinkelshtein1972}.  In particular, in the quantum tunneling creation of quantized vortices in superfluids \cite{Volovik1972} and superconductors \cite{Blatter1994}, the dynamically conjugate variables are the area of the vortex ring and its coordinate along the normal to the ring. This approach provides the correct semiclassical tunneling exponent without consideration of the details of the structure of the object on the microscopic level. In the same manner, it looks reasonable that in case of tunneling from the black hole to the white hole, described by the collective variables $K$ and $A$, the microscopic processes are not important.

In the case of Hawking radiation, the tunneling of particle from inside to outside the horizon is obtained by evaluating the action of the particle \cite{Kraus1997,Berezin1999,Volovik1999,Wilczek2000,Volovik2009}. The tunneling rate is determined by imaginary part of the action and can be obtained using the path in the complex plane. The calculation of the tunneling exponent demonstrates that it is proportional to the change of the black hole entropy after radiation of a particle, $p \propto e^{\Delta S_{\text{BH}}}$, see Refs. \cite{Kraus1997,Berezin1999,Wilczek2000}. The same result can be also obtained using the conjugate variables $K$ and $A$  in the Euclidean action, and the path $\int A(K) dK$ at fixed $M$ with real $K$. The direct connection between the Hawking radiation and black hole tunneling will be discussed in Sec. \ref{WHentropy}.

As in the case of the semiclassical consideration of the Hawking radiation in terms of the quantum tunneling \cite{Volovik1999,Wilczek2000,Volovik2009},
we shall use the Painleve-Gullstrand coordinate system \cite{Painleve,Gullstrand} with the metric:
 \begin{equation}
ds^2= - dt^2(1-{\bf v}^2) - 2dt\, d{\bf r}\cdot {\bf v} + d{\bf r}^2 \,.
\label{PGmetric}
\end{equation}
Here the vector $v_i({\bf r})=g_{0i}({\bf r})$ is the velocity of the free-falling observer, which crosses the horizon. 
In condensed matter, the analogs of the black hole and white hole horizons described by this metric (known as acoustic metric  \cite{Unruh1981}) can be reproduced in the Dirac and Weyl topological semimetals,
where the horizon takes place on the boundary between different types of Dirac or Weyl materials \cite{Volovik2016,Kuang2017,Wilczek2020,Nissinen2017}.

For the Schwartzschild black hole one has
\begin{equation}
{\bf v} ({\bf r})=\mp \hat{\bf r} \sqrt{\frac{r_\text{H}}{r}}=\mp \hat{\bf r} \sqrt{\frac{M}{2rK}}\,,
\label{velocity}
\end{equation}
where $r_\text{H}$ is the radius of the horizon; the minus sign corresponds to the black hole and the plus sign describes the white hole \cite{Barcelo2014}.
Note that in the theory with the variable gravitational coupling, the sign changes at the singularity $K=\infty$ (or at $G=0$), when 
the black hole shrinks to a point and then expands as a white hole. In terms of variable $K$, the point $K=\infty$  serves as the branch point, where the velocity of the freely falling observer changes sign. 

It is important that the vector ${\bf v}$, which is normal to the surface of the horizon, is the velocity of the free falling observer, who crosses the horizon. For the observer, who crosses the black hole horizon from outside, and for the observer, who crosses the white hole horizon from inside, these two directions are opposite. For these two observers the area of the horizons has different sign. This means that at the branch point of the trajectory the horizon area $A$ changes sign: it crosses zero at $K=\infty$ and  becomes negative on the white-hole side of the process, $A \rightarrow -A$. This also could mean that due to connection between the area and entropy the white hole may have negative entropy, which we discuss in Sec.\ref{WHentropy}.

The quantum tunneling exponent is usually determined by the imaginary part of the action on the trajectory, which transforms the black hole to white hole. In terms of Euclidean action one has:
 \begin{equation}
p\propto \exp{\left(-I_{\text {BH} \rightarrow \text{WH}}\right)} \,\,,\,\, I_{\text{BH} \rightarrow \text{WH}} =  \int_C A(K')dK' \,.
\label{TunnelingExponent}
\end{equation}
Here the semiclassical trajectory $C$ is at $M=\rm{const}$, and thus $A(K')=\pm \pi M^2/(K')^2$. Along this trajectory the variable $K'$ changes from $K$ to the branch point at $K'=\infty$, and then from  $K'=\infty$ to $K'=K$ along the other  branch, where the area $ A(K')<0$.  
The integral gives the tunneling exponent of the transition to the white hole 
\begin{equation}
I_{\text{BH} \rightarrow \text{WH}}=  \,  2\pi M^2     \int_K^\infty \frac{dK'}{K'^2}= 2 \pi \frac{M^2}{K}  \,,
\label{BohrQuantization}
\end{equation}
and the transition probability is:
 \begin{equation}
p\propto \exp{\left(-2\pi M^2/K\right)}=\exp{\left(-2S_\text{BH}\right)}
\,.
\label{tunneling}
\end{equation}

\section{White hole entropy and temperature}
\label{WHentropy}

The result (\ref{tunneling}) can be also obtained using the Hawking radiation from the black hole, where the tunneling exponent is proportional
to $e^{\Delta S_{\text{BH}}}$, see Refs. \cite{Kraus1997,Berezin1999,Wilczek2000}. Let us consider the process, in which the particle escapes the black hole by quantum tunneling and then it tunnels to the white hole through the white hole horizon. This process occurs at the fixed total mass $M$. The tunneling exponent for this process  to occur is $e^{(\Delta S_{\text{BH}}+\Delta S_{\text{WH}})}$. 
Summation of all the processes of the tunneling of matter from the black hole to the formed white hole gives finally Eq.(\ref{tunneling}):
\begin{equation}
p\propto e^{\sum(\Delta S_{\text{BH}}+\Delta S_{\text{WH}})}=e^{2\sum\Delta S_{\text{BH}}}=\exp{\left(-2S_\text{BH}\right)}
\,.
\label{tunneling2}
\end{equation}
Here we took into account the (anti)symmetry in the dynamics of black and white holes in the process of quantum tunneling,
$\sum\Delta S_{\text{BH}}=\sum \Delta S_{\text{WH}}$.

As in the case of quantum tunneling in the Hawking radiation process, the probability in Eq.(\ref{tunneling}) has the thermodynamic meaning as thermodynamic fluctuation $e^{\Delta S}$ \cite{Landau_Lifshitz}. In our case the total change of the entropy in the process of the tunneling from black to white hole is
$\Delta S=S_\text{WH}-S_\text{BH}$. According to Eq.(\ref{tunneling2}) this change is equal to $-2S_\text{BH}$. As a result, one obtains that the entropy of the white hole is equal with the opposite sign to the entropy of the black hole with the same mass 
\begin{equation}
S_\text{WH}(M)=-S_\text{BH}(M)
\,.
\label{fluctuation}
\end{equation}
This means that the white hole, which is obtained by quantum tunneling from the black hole and thus has the same mass $M$ as the black hole, has the negative temperature 
 $T_\text{WH}=- T_\text{BH}$ and the negative area $A_\text{WH}=- A_\text{BH}$, which together produce the negative entropy $S_\text{WH}=- S_\text{BH}$.
 
The negative temperature is a well defined quantity in condensed matter. It typically takes place in the subsystem of nuclear spins, where the energy is restricted from above. Different  thermodynamic phase transitions occurring at $T<0$ have been experimentally observed in magnetic systems, see e.g. Ref. \cite{Hakonen1992}. The negative energy states are unstable both before and after the magnetic phase transition. In principle, the state with $T<0$ is hotter than the state $T>0$, since the heat flows from the negative- to the positive-temperature system. If the black hole and white hole are in some contact,  the heat will excape from the white hole and will be absorbed by the black hole.

The transition from positive to negative $T$ occurs on the path $A(K)dK$ via the point $K=\infty$, where $T(K)=\infty$. The transition via infinite temperature is the analytic route to the thermodynamically unstable states with negative $T$, see e.g. Refs. \cite{Volovik2019,Volovik2019b}, where the transition to anti-spacetime \cite{Christodoulou2012,Boyle2018,Boyle2018b} has been considered. In spin systems the $T<0$ state is typically obtained by reversing the magnetic field, which looks like crossing $T=0$.

In the black hole physics the negative temperature has been discussed for the inner horizon of the Kerr and charged black holes, see Ref.\cite{Gibbons2018} and references therein. The entropy of the inner horizon has been considered as positive.
 However, the arguments in Ref.\cite{Gibbons2018} do not exclude the possibility of the opposite situation, when the temperature of the inner horizon is positive, while the entropy is negative: the equation $T_+S_++T_-S_-=0$ remains valid.

The discussed transition from the black hole to the white hole with the same mass $M$ is not the thermodynamic transition. It is the quantum process of tunneling between the two quantum states. It is one of many routes of the black hole evaporation, including formation of small white hole on the late stage of the decay \cite{Barcelo2014,Barcelo2017,Rovelli2018,Rovelli2019,Rovelli2018b,Uzan2020,Uzan2020b,Uzan2020c,Bodendorfer2019}.
The uniqueness of this route together with hidden information and (anti)symmetry between the black and white holes is the possible origin of the negative entropy of the white hole.

\section{Discussion}

We considered the dynamics and thermodynamics of the black hole in case of the varying gravitational coupling $K$. The gravitational coupling $K$ serves as the thermodynamic variable, which is thermodynamic conjugate to $A_\text{BH}T_\text{BH}$, where 
$A_\text{BH}$ and $T_\text{BH}$ are correspondingly the area of the black hole horizon and Hawking temperature.
The corresponding first law of the black hole is modified, see Eq.(\ref{dS}). The corresponding adiabatic invariant is  the entropy $S_\text{BH}=KA_\text{BH}$, and it is this invariant which should be quantized, if the Bekenstein conjecture is correct.
This is in agreement with  observation of Ted Jacobson \cite{Jacobson1994}, that it is the entropy that does not change under renormalization of $K$, rather than the area.  This suggests the alternative quantization scheme for the black hole.  While $K$ and $A_\text{BH}$ are dimensionful  and cannot be quantized, the entropy is dimensionless and thus can be quantized in terms of some  fundamental numbers.   

On the quantum level the dynamically conjugate variables of the black hole physics are $K$ and $A_\text{BH}$. This allows us to consider
the transition from the black hole to the white hole as quantum tunneling in the semiclassical approximation, which is valid when the action is large. The classical trajectory of the black hole crosses the branch point at $K=\infty$, and then continues along the other branch, where the area $ A(K)<0$, which corresponds to the white hole. The obtained tunneling exponent $\exp(-2S_\text{BH})$ demonstrates that the transition can be considered as thermodynamic fluctuation, if the entropy of the white hole (with the same mass $M$  as the black hole) is negative, $S_\text{WH}=- S_\text{BH}$. The latter also suggests that this white hole has negative temperature $T_\text{WH}=- T_\text{BH}$.

  {\bf Acknowledgements}. This work has been supported by the European Research Council (ERC) under the European Union's Horizon 2020 research and innovation programme (Grant Agreement No. 694248). I thank J. Barrow and A. Starobinsky for the criticism and H. Terazawa for correspondence.

\end{document}